\documentclass[aps,superscriptaddress,floatfix,showpacs]{revtex4}

\usepackage{amsmath}
\usepackage{mathrsfs}
\usepackage{bm}

\newcommand{\ppi}{\bm{\mathnormal{\Pi}}}
\newcommand{\arctanh}{\text{arctanh}}

\begin{document}

\title{Dynamics of radiation due to vacuum nonlinearities}

\author{Mattias Marklund}
\email{marklund@elmagn.chalmers.se}
\affiliation{Department of Electromagnetics, Chalmers University of
Technology, SE--412 96 G\"oteborg, Sweden}

\author{Gert Brodin}
\author{Lennart Stenflo}
\affiliation{Department of Physics, Ume{\aa} University,
SE--901 87 Ume{\aa}, Sweden}

\author{Padma K.\ Shukla}
\affiliation{Institut f\"ur Theoretische Physik IV, Fakult\"at f\"ur
  Physik und Astronomie, Ruhr-Universit\"at
  Bochum, D--447 80 Bochum, Germany}
\affiliation{Department of Physics, Ume{\aa} University, SE--901 87
  Ume{\aa}, Sweden}

\date{\today}

\begin{abstract}
In quantum electrodynamics, photon--photon scattering can be the result of
the exchange of virtual electron--positron pairs. This gives
rise to a non-trivial dispersion relation for a single photon moving
on a background of electromagnetic fields. Knowledge of the dispersion
relation can be transferred, using standard methods, into new insights in the
dynamical equations for the photons. Effectively, those equations will
contain different types of self-interaction terms, depending on
whether the photons are coherent or not. It is shown that coherent
photons are governed by a nonlinear Schr\"odinger type
equation, such that the
self-interaction terms vanish in the limit of parallel propagating
waves. For incoherent photons, a set of fluid equations can
determine the evolution of the corresponding radiation
gas. In the case of a self-interacting radiation fluid, it is shown
that Landau damping can occur.
\end{abstract}

\pacs{12.20.Ds, 95.30.Cq}

\maketitle


\section{Introduction}


According to QED, the non-classical phenomenon of photon--photon scattering
can take place due to the exchange of virtual electron--positron pairs. This
is a second order effect (in terms of the fine structure constant $\alpha
\equiv e^{2}/4\pi \varepsilon _{0}\hbar c\approx 1/137$), which in standard
notations can be deduced from the Euler--Heisenberg Lagrangian
density \cite{Heisenberg-Euler,Schwinger}
\begin{equation}
{\mathscr{L}}=\varepsilon_0{\mathscr{F}} + \varepsilon_0^2\kappa
(4{\mathscr{F}}^{2}+7{\mathscr{G}}^{2})\ ,  \label{eq:lagrangian}
\end{equation}
where $\kappa \equiv 2\alpha ^{2}\hbar ^{3}/45m_{e}^{4}c^{5}
$, ${\mathscr{F}}\equiv \frac{1}{2}(E^{2}-c^{2}B^{2})$,
${\mathscr{G}}\equiv
c{\bf E}\cdot {\bf B}$,  $e$ is the electron charge,
$c$ the velocity of
light, $2\pi \hbar $ the Planck constant and $m_{e}$ the electron
mass.
The Lagrangian (\ref{eq:lagrangian}) is
valid as long as there is no pair creation and the field strength is
smaller than the critical field, i.e.
\begin{equation}
  \omega \ll m_ec^2/\hbar \,\,\text{ and }\,\, |\bm{E}| \ll
  E_{\text{crit}} \equiv
  m_ec^2/e\lambda_c
  \label{eq:constraint}
\end{equation}
respectively. Here $\lambda_c$ is the
Compton wave length, and $E_{\text{crit}} \simeq
10^{18}\,\mathrm{V}/\mathrm{m}$.
The latter terms in (\ref{eq:lagrangian}) represent the effects of vacuum
polarisation and magnetisation. We note that ${\mathscr{F}}={\mathscr{G}}=0$
in the limit of parallel propagating waves. It is therefore necessary to use
other waves in order to obtain an effect from the QED corrections. Several
attempts have been presented in the literature
\cite{valluri,Rozanov1998,Rozanov1993,Alexandrov,segev,Ding,%
Brodin-Marklund-Stenflo}, where Refs.\ \cite{valluri,Rozanov1998,%
Rozanov1993,Alexandrov} mainly focused on principal
issues, whereas the experimental possibilities for detection have been
discussed in Refs.\ \cite{segev,Ding,Brodin-Marklund-Stenflo}.

The non-trivial
propagation of photons in strong background
electromagnetic fields, due to effects of nonlinear electrodynamics,
has been considered in a number of papers (see, e.g., Refs.\
\cite{Bialynicka-Birula,Boillat} and
references therein). Their main focus was on the
interesting effects of photon splitting and birefringence in
vacuum. However, Thoma \cite{Thoma} investigated the interaction of
photons with
a photon gas, using the real-time formalism, and calculated the
corresponding change in the speed of light due to the Cosmic Microwave
Background (CMB).

In the present paper, we
derive an evolution equation for an ensemble of electromagnetic
pulses. Moreover, the methods of
radiation hydrodynamics \cite{Mihalas} are combined with the QED
theory for
photon--photon scattering, and a system of equations (c.f. \cite{Karpman}) is
obtained, where the radiation pressure of the pulse
will act as a driver of acoustic waves in the photon gas.


\section{Equations and examples}


In a medium with polarisation ${\bf P}$ and magnetisation ${\bf M}$ the
general wave equations for ${\bf E}$ and ${\bf B}$ are
\begin{subequations}
\begin{equation}
\frac{1}{c^2}\frac{\partial^2{\bf E}}{\partial t^2} - \nabla^2{\bf E} =
-\mu_0\left[ \frac{\partial^2{\bf P}}{\partial t^2} + c^2\nabla(\nabla\cdot%
{\bf P}) + \frac{\partial}{\partial t}(\nabla\times{\bf M)} \right] \ ,
\label{WaveE}
\end{equation}
and
\begin{equation}
\frac{1}{c^2}\frac{\partial^2{\bf B}}{\partial t^2} - \nabla^2{\bf B} = \mu_0%
\left[ \nabla\times(\nabla\times{\bf M}) + \frac{\partial}{\partial t}%
(\nabla\times{\bf P)} \right] \ .  \label{WaveB}
\end{equation}
\end{subequations}
Using the Lagrangian (\ref{eq:lagrangian}), one can show that due to
photon-photon scattering induced by the exchange of virtual
electron-positron pairs, the vacuum will effectively be polarised and
magnetised,
and the polarisation and magnetisation will be nonlinear to cubic
order in the electromagnetic field (see, e.g., Ref.\
\cite{segev}). These equations can be used to analyse the properties
of coherent radiation, and it is straightforward to show that the
non-linearities vanish in the case of co-propagating plane waves. On
the other hand, introducing dispersion into the propagation, by the use
of wave guides, will give
non-trivial polarisations and magnetisation, and this has been used as
a suggested means of detection of photon--photon scattering
\cite{Brodin-Marklund-Stenflo}, and to form light bullets in vacuum
\cite{Brodin-etal}.

On the other hand, as shown in Ref.\ \cite{Bialynicka-Birula}, one can
derive the dispersion relation for a single photon moving on a given
background of electromagnetic fields, using first principles. The
result is (see also Ref.~\cite{Boillat} and references
therein)
\begin{equation}\label{eq:dispersionrelation}
  \omega(\bm{k}, \bm{E}, \bm{B}) = c|\bm{k}|\left( 1 -
  \tfrac{1}{2}\lambda|\bm{Q}|^2 \right) .
\end{equation}
where
\begin{eqnarray}
  |\bm{Q}|^2 = \varepsilon_0\left[ E^2 + c^2B^2
   -
   (\hat{\bm{k}}\cdot\bm{E})^2 -
   c^2(\hat{\bm{k}}\cdot\bm{B})^2 -
   2c\hat{\bm{k}}\cdot(\bm{E}\times\bm{B})\right] ,
\label{eq:Q2}
\end{eqnarray}
and $\lambda = \lambda_{\pm}$, where $\lambda_+ = 14\kappa$ and
$\lambda_-
= 8\kappa$ for the two different polarisation states of the
photon. Furthermore, $\hat{\bm{k}} \equiv \bm{k}/k$.
The approximation $\lambda|\bm{Q}|^2 \ll 1$ has been used.
The background electric and magnetic fields are
denoted by $\bm{E}$ and $\bm{B}$, respectively.

From expressions (\ref{eq:dispersionrelation}) and (\ref{eq:Q2}) we
can derive essentially different dispersion relations, depending
on whether or not the background is coherent or incoherent.

To start with, suppose a photon with wave vector $\bm{k}$ is moving on
a background of photons close to thermal equilibrium. The
background electromagnetic fields then satisfy $\langle\bm{E}\rangle =
\langle\bm{B}\rangle = 0$, $\langle E_iE_j \rangle = \langle E^2
\rangle\delta_{ij}/3$, $\langle B_iB_j \rangle = \langle B^2
\rangle\delta_{ij}/3$, where the angular brackets denote ensemble average.
From (\ref{eq:Q2}) we then obtain
\begin{equation}
  \label{eq:Q2gas}
  |\bm{Q}_g|^2 = \tfrac{4}{3}\mathscr{E}_g ,
\end{equation}
where $\mathscr{E} = \varepsilon_0(\langle E^2 \rangle + c^2\langle
B^2 \rangle)/2$ is the energy density of the radiation gas.

On the other hand, if the photon propagates on a plane wave
background, such that $\bm{E} =
E\hat{\bm{e}}$, and $\bm{B} =
E\hat{\bm{k}}_p\times\hat{\bm{e}}/c$, where
$\hat{\bm{e}}$ is the unit electric vector and $\bm{k}_p$ is the wave
vector of the background field, we find that
\begin{equation}\label{eq:Q2pulse}
  |\bm{Q}|^2 = \left[ 2 - 2(\hat{\bm{k}}\cdot\hat{\bm{k}}_p) -
   (\hat{\bm{k}}\cdot\hat{\bm{e}})^2 -
   (\hat{\bm{k}}\cdot(\hat{\bm{k}}_p\times\hat{\bm{e}}))^2
   \right]\varepsilon_0|E|^2  .
\end{equation}
If $\bm{k} = \bm{k}_p$, the interaction term $|\bm{Q}|^2$ vanishes
identically, consistent with the previous statement that a plane wave
pulse can not self-interact. On the other hand, if the single photon
belongs to a random ensemble of photons, i.e., a photon gas close to
thermal equilibrium, we have
$\langle\hat{\bm{k}}\rangle = 0$ and $\langle \hat{k}_i\hat{k}_j
\rangle \delta_{ij}/3$. The expression (\ref{eq:Q2pulse}) then reduces to
\begin{equation}
|\bm{Q}_p|^2 =
   \tfrac{4}{3}\mathscr{E}_p ,
\end{equation}
where $\mathscr{E}_p = \varepsilon_0|E|^2$ is the energy density of
the background electromagnetic pulse.

Using the dispersion relation above, different scenarios can be
analysed.

\subsection{An ensemble of pulses}

For a plane wave pulse moving on a background of photons, randomly
distributed in momentum space, the relation (\ref{eq:Q2gas}) holds. It
is then straightforward to derive, by using
the eikonal representation and the WKBJ approximation \cite{Karpman},
the Schr\"odinger equation \cite{Marklund-Brodin-Stenflo}
\begin{equation}\label{eq:nlse}
  i\left( \frac{\partial}{\partial t} +
  c{\hat{\bm{k}}}_0\cdot\nabla
  \right)E +
  \frac{c}{2k_0}\left[\nabla^2 - ({\hat{\bm{k}}}_0\cdot\nabla)^2\right]
  E +
  \frac{2}{3}\lambda ck_0\mathscr{E}_gE = 0 ,
\end{equation}
where $\bm{k}_0$ is the linear vacuum wave number of the pulse. Suppose now
that we have an ensemble of pulses. We decompose the
electric field according to $E =
\langle E\rangle + \widetilde{E}$,
where $\langle\widetilde{E}\rangle = 0$. Taking the average of
Eq.~(\ref{eq:nlse}), we obtain
\begin{eqnarray}
  i\left( \frac{\langle k_0\rangle}{c}\frac{\partial}{\partial t} +
  \langle{k}_{0i}\rangle\frac{\partial}{\partial x_i}
  \right)\langle E\rangle + \frac{1}{2}\left(
  \delta_{ij} - \langle\hat{k}_{0i}\hat{k}_{0j} \rangle
  \right)\frac{\partial^2\langle E\rangle}{\partial x_i
  \partial x_j} +  \frac{2}{3}\lambda\langle
  k_0^2\rangle\mathscr{E}\langle E\rangle
   = \left\langle\!\!
  -i\frac{k_0}{c}\frac{\partial\widetilde{E}}{\partial
  t}
  - i
  {k}_{0i}\frac{\partial\widetilde{E}}{\partial x_i}
  +
  \frac{1}{2}\hat{k}_{0i}\hat{k}_{0j}
  \frac{\partial^2\widetilde{E}}{\partial
  x_i \partial x_j}\!\!\right\rangle\!  ,
\label{eq:expandednlse}
\end{eqnarray}
where we have employed Einstein's summation convention. In general, we
can subtract Eq.~(\ref{eq:expandednlse}) from (\ref{eq:nlse}) in order to
obtain a similar equation for $\widetilde{E}$. This
would then constitute a general coupled system between the fluctuations
and the average wave packet. We will not pursue this issue further, however.

We now write $\hat{\bm{k}}_0 =
\langle\hat{\bm{k}}_0\rangle + \tilde{\hat{\bm{k}}}_0$ with
$\langle\tilde{\hat{\bm{k}}}_0\rangle = 0$, which implies
$\langle\hat{k}_{0i}\hat{k}_{0j}\rangle =
  \langle\hat{k}_{0i}\rangle\langle\hat{k}_{0j}\rangle + \langle
  \tilde{\hat{k}}_{0i}\tilde{\hat{k}}_{0j}\rangle$.
If $\tilde{\hat{\bm{k}}}_0$ is approximately isotropically
distributed, we then have
$\langle\tilde{\hat{k}}_{0i}\tilde{\hat{k}}_{0j}\rangle =
  \tfrac{1}{3}\langle|\tilde{\hat{\bm{k}}}_0|^2\rangle\delta_{ij}$.
Furthermore, since $|\hat{\bm{k}}_0|^2 = 1$, it follows that
$0 \leq\langle|\tilde{\hat{\bm{k}}}_0|^2\rangle = 1 -
|\langle\hat{\bm{k}}_0\rangle|^2 \leq 1$. Thus, 
\begin{equation}
  \langle\hat{k}_{0i}\hat{k}_{0j}\rangle =
  \langle\hat{k}_{0i}\rangle\langle\hat{k}_{0j}\rangle +
  \tfrac{1}{3}\left( 1 - |\langle\hat{\bm{k}}_0\rangle|^2
  \right)\delta_{ij} .
\label{eq:direction2}
\end{equation}

In the case of small random fluctuations in the field
amplitude, i.e., $|\widetilde{E}| \ll
|\langle E\rangle|$, we neglect the right hand side
of Eq.~(\ref{eq:expandednlse}). Suppose furthermore that the averaged
wave vector, its direction, and its amplitude all change slowly over
time and space compared to the amplitude
$\langle E\rangle$. Moreover, for a self-interacting
pulse ensemble, we see that $\mathscr{E} =
\varepsilon_0\langle|E|^2\rangle \approx
\varepsilon_0|\langle E\rangle|^2$. Using the relation
(\ref{eq:direction2}), we thus obtain the dynamical equation
\begin{equation}
  i\left( \frac{\partial}{\partial t} + \frac{c}{\langle
  k_0\rangle}\langle\bm{k}_0\rangle\cdot\nabla
  \right)\langle E\rangle + \frac{c}{2\langle
  k_0\rangle}\left[
  \frac{2}{3}\left( 1 + \frac{1}{2}|\langle\hat{\bm{k}}_0\rangle|^2
  \right)\nabla_{\perp}^2 + \frac{2}{3}\left( 1 -
  |\langle\hat{\bm{k}}_0\rangle|^2 \right)\nabla_{||}^2
  \right]\langle E\rangle + \frac{2}{3}\frac{\lambda c\varepsilon_0\langle
  k_0^2\rangle}{\langle
  k_0\rangle}\,|\langle E\rangle|^2\langle E\rangle = 0
  ,
\label{eq:approxnlse}
\end{equation}
where  $\nabla_{\perp}^2 = \nabla^2 - \nabla_{||}^2$ and
$\nabla_{||}^2 = (\langle
\hat{k}_{0i}\rangle\langle \hat{k}_{0j}\rangle/|\langle
\hat{\bm{k}}_{0}\rangle|^2)\partial_i\partial_j$. Equation
(\ref{eq:approxnlse}) therefore describes the dynamics of the
non-fluctuating part of an ensemble of pulses, when the nonlinear
self-interaction due to photon--photon scattering is taken into
account.

\subsection{The kinetic theory of
  non-linear photons}

In general, from our dispersion relation one can formulate the
single-particle dynamics in terms of the Hamiltonian ray equations.
For a dispersion relation $\omega = ck[1 - (\lambda/2)|\bm{Q}|^2]$,
where $|\bm{Q}|^2$ is assumed to be independent of $\bm{k}$, we have the
Hamiltonian ray equations
\begin{subequations}
\begin{eqnarray}
  \dot{\bm{r}} &=& \frac{\partial\omega}{\partial\bm{k}} =
  c\left(1 - \frac{1}{2}\lambda|\bm{Q}|^2\right)\hat{\bm{k}}  ,
  \label{eq:groupvelocity} \\
  \dot{\bm{k}} &=& -\frac{\partial\omega}{\partial\bm{r}}
    =
    \frac{1}{2}\lambda ck\frac{\partial|\bm{Q}|^2}{\partial\bm{r}}
    ,
\label{eq:kdot}
\end{eqnarray}
\end{subequations}
where $\dot{\bm{r}}$ is the group velocity of the photon,
$\dot{\bm{k}}$ is the force on a photon, and the dot denotes time
derivative.

The equation for the collective interaction of photons can
then be formulated as~\cite{Marklund-Brodin-Stenflo,Mendonca}
\begin{equation}\label{eq:kinetic}
  \frac{\partial N}{\partial t} +
  c\left(1 - \frac{1}{2}\lambda|\bm{Q}|^2\right)\hat{\bm{k}} \cdot
  \frac{\partial N}{\partial\bm{r}} +
  \frac{1}{2}\lambda ck\frac{\partial|\bm{Q}|^2}{\partial\bm{r}} \cdot
  \frac{\partial N}{\partial\bm{k}} = 0 .
\end{equation}
where the distribution function $N = N(\bm{k}, \bm{r}, t)$
has been normalised such that the
number density is $n(\bm{r}, t) = \int N(\bm{k}, \bm{r},t)\,d\bm{k}$.

From Eq.\ (\ref{eq:kinetic}) a hierarchy of fluid equations can be
built  \cite{Marklund-Brodin-Stenflo}, of which the two first are
the energy conservation equation
\begin{subequations}
\begin{equation}\label{eq:energy}
  \frac{\partial\mathscr{E}_g}{\partial t} +
  \nabla\cdot\left( \mathscr{E}_g\bm{u} + \bm{q} \right) =
  -\frac{1}{2}\lambda\mathscr{E}\frac{\partial|\bm{Q}|^2}{\partial
  t} ,
\end{equation}
and the momentum conservation equation
\begin{eqnarray}\label{eq:momentum}
  \frac{\partial{\ppi}}{\partial t} + \nabla\cdot\Big[
  \bm{u}\otimes{\ppi} + \bm{\mathsf{P}} \Big] =
  \frac{1}{2}\lambda\mathscr{E}\nabla|\bm{Q}|^2 ,
\end{eqnarray}
\label{eq:comoving}
\end{subequations}
respectively. Here
\begin{equation}\label{eq:energy-def}
  \mathscr{E}_g(\bm{r}, t) = \int\hbar\omega N\,d\bm{k}
\end{equation}
is the energy density, $\bm{q}(\bm{r}, t) =
\int{\hbar\omega\bm{w}} N\,d\bm{k}$
the energy flux, and
we have made the split $\dot{\bm{r}} = \bm{u} + \bm{w}$, where
$\langle{\bm{w}}\rangle = 0$.
Furthermore, ${\ppi} = \int{\hbar\bm{k}}N\,d\bm{k}$
is the momentum density, and
$\bm{\mathsf{P}} = \int{\bm{w}\otimes(\hbar\bm{k})}N\,d\bm{k}$
is the pressure tensor.

\subsection{The interaction between a
  radiation gas and an electromagnetic pulse}

From Eqs.\ (\ref{eq:Q2gas}) and (\ref{eq:Q2pulse}) it is clear that
even though the interaction of a pulse with itself will vanish, this
is not the case if the pulse moves on a background of incoherent
photons, i.e., a radiation gas. The problem of an interacting pulse
and an incoherent radiation gas via photon--photon scattering  was
investigated in Ref.\ \cite{Marklund-Brodin-Stenflo}, where a coupled
system of equations was derived \cite{Karpman}.

\subsection{The self-interacting
  radiation gas}

From (\ref{eq:Q2gas}) it follows that a kinetic photon gas may
interact with itself via a ponderomotive force $\propto
\nabla\mathscr{E}_g$. Using Eqs.\ (\ref{eq:Q2gas}) and
(\ref{eq:kinetic}), the kinetic equation for
this type of gas is
\begin{equation}
  \frac{\partial N}{\partial t} + c\left( 1 -
  \frac{2}{3}\lambda\mathscr{E} \right)\hat{\bm{k}}\cdot\frac{\partial
  N}{\partial\bm{r}} +
  \frac{2}{3}ck\lambda\frac{\partial\mathscr{E}}{\partial\bm{r}} \cdot
  \frac{\partial N}{\partial\bm{k}} = 0 .
\label{eq:kinetic-new}
\end{equation}

Assuming $N(\bm{k}, \bm{r}, t) = N_0(\bm{k}) + N_1(\bm{k})
\exp[i(\bm{K}\cdot\bm{r} - \Omega t)]$, where $N_1 \ll N_0$ and
$\mathscr{E}(\bm{r},t) = \mathscr{E}_0 +
\mathscr{E}_1\exp[i(\bm{K}\cdot\bm{r} - \Omega t)]$, where $\mathscr{E}_1
\ll \mathscr{E}_0$, we
linearise Eqs.\ (\ref{eq:kinetic-new}) and (\ref{eq:energy-def}) to
obtain the dispersion
relation
\begin{equation}
  1 = \frac{2}{3}\lambda\hbar c^2\int \frac{k^2\,\bm{K}\cdot(\partial
  N_0/\partial\bm{k})}{\Omega - c\bm{K}\cdot\hat{\bm{k}}}\, d\bm{k} .
\label{eq:dispersion-relation}
\end{equation}

If the background distribution is isotropic, i.e., $N_0(\bm{k}) =
N_0(k)$, and introducing spherical coordinates in $\bm{k}$-space,
we rewrite Eq.\ (\ref{eq:dispersion-relation}) as
\begin{equation}
  1 = \frac{4\pi}{3}\lambda\hbar c \int_{-1}^1\frac{\alpha}{\beta -
  \alpha}\,d\alpha\int_{0}^{\infty} k^4\frac{dN_0}{dk}\,dk
\end{equation}
where $\beta = \Omega/cK$ and $\alpha = \hat{\bm{k}}\cdot\bm{K}/K$.
Performing the angular integration in $\alpha$, for $\beta < 1$, we
obtain
\begin{equation}
  1 = \frac{8\pi}{3}\lambda\hbar c \left[ -1 + \beta\,\arctanh\,(\beta) -
  \pi i\beta \right] \int_{0}^{\infty}
  k^4\frac{dN_0}{dk}\,dk .
  \label{eq:dispersion-relation2}
\end{equation}
We now assume that we have a Gaussian
background distribution
\begin{equation}
  N_0(k) = \frac{\mathscr{E}_0}{8\pi a^4 c\hbar}
   \exp\left[ -\frac{k^2}{2a^2}
  \right] ,
\end{equation}
where $a$ is the width of the distribution. The dispersion relation
(\ref{eq:dispersion-relation2}) is then
\begin{equation}
  1 = \frac{8}{3}\lambda\mathscr{E}_0 \left[ -1 + \beta\,\arctanh\,(\beta) -
  \pi i\beta \right] .
  \label{eq:dispersion-relation3}
\end{equation}


\section{Conclusion}


In the present paper we have considered the interaction, due to the quantum
electrodynamic photon-photon scattering, of electromagnetic
waves and incoherent photons. For this purpose we have applied the
dispersion relation derived in Ref.\ \cite{Bialynicka-Birula}. The
dynamical equations for an ensemble of pulses has been derived, and a
set of equations describing a radiation fluid has been presented. An
expression showing Landau damping within the self-interacting
radiation gas has been derived.



\end{document}